\documentclass{Interspeech}

\interspeechcameraready 

\title{Revisiting the Privacy of Low-Frequency Speech Signals: Exploring Resampling Methods, Evaluation Scenarios, and Speaker Characteristics}
\author[affiliation={1,2}]{Jule}{Pohlhausen}
\author[affiliation={1}]{Joerg}{Bitzer}

\affiliation{Institute of Hearing Technology and Audiology}{Jade University of Applied Sciences, Oldenburg}{Germany}
\affiliation{Department of Medical Physics and Acoustics}{Carl von Ossietzky Universität Oldenburg}{Germany}

\email{jule.pohlhausen@jade-hs.de}


\usepackage{comment}
\usepackage{caption}
\usepackage{subcaption}
\usepackage[export]{adjustbox}
\usepackage{csquotes}

\begin{document}

\maketitle

\begin{abstract}
While audio recordings in real life provide insights into social dynamics and conversational behavior, they also raise concerns about the privacy of personal, sensitive data.
This article explores the effectiveness of restricting recordings to low-frequency audio to protect spoken content. For resampling the audio signals to different sampling rates, we compare the effect of employing anti-aliasing filtering.
Privacy enhancement is measured by an increased word error rate of automatic speech recognition models. The impact on utility performance is measured with voice activity detection models. 
Our experimental results show that for clean recordings, models trained with a sampling rate of up to \SI{800}{\hertz} transcribe the majority of words correctly. For both models, we analyzed the impact of the speaker's sex and pitch, and we demonstrated that missing anti-aliasing filters more strongly compromise speech privacy.
\end{abstract}

\section{Introduction}
Audio recordings are ubiquitous in our everyday lives, with applications ranging from monitoring noise pollution \cite{maisonneuve2009citizen} or animal biodiversity \cite{Darras2016} to detecting anomalous audio events in surveillance systems \cite{valenzise2007scream} or analyzing social behavior \cite{mehl2017ear}.
This article focuses on the latter application and the challenges of recording speech signals.
Speech recordings can be processed to obtain personal, sensitive information such as full names, political opinions, and the speaker's gender or health status \cite{nautsch2019speech_privacy}.
As a result, the European general data protection regulation (EU GDPR) states that these recordings require privacy preservation and the protection of speech data.

Most existing solutions for privacy-preserving recording rely on wearable recording systems, which tend to limit computational resources and memory.
To address this, various privacy-preserving safeguards have been proposed, including the recording of audio snippets \cite{mehl2017ear} or acoustic features \cite{parthasarathi2011privacy, bitzer2016privacy}, employing encryption or distributed
learning \cite{backstrom2023privacy}.
Another approach is to obfuscate speech content, e.g., with sound shredding \cite{kumar2015sound}, slicing \cite{maouche2022enhancing}, or temporal and spectral smoothing \cite{Pohlhausen2025csl}.
A simpler option is to reduce the sampling rate, which limits the frequency content and also reduces data storage requirements.
For instance, \cite{raman2022conflab} developed a wearable multi-sensor recorder that stores low-frequency audio with a sampling rate of \SI{1250}{\hertz} and released the ConfLab dataset of social conversations among 48 conference attendees.
However, the effectiveness of concealing linguistic content was only evaluated by employing an open-source, pre-trained automatic speech recognition (ASR) model \cite{liu2024lowfreq} and additional bandwidth extension. 
Similarly, \cite{Cohen2019urban} reduced the sampling rate to \SI{500}{\hertz} to anonymize voices recorded in public spaces, but their privacy evaluation also relied on pre-trained attacker models.
We hypothesize that this evaluation scenario overestimates the protection of linguistic contents, as the ASR model was not specifically trained or finetuned on low-frequency audio (see Section~\ref{sec:attack} for details).
In fact, \cite{Pohlhausen2025csl, boovaraghavan2024kirigami} demonstrated the stronger attack capabilities of ASR models fine-tuned with sampling rates of \SI{1250}{\hertz} and \SI{1}{\kilo\hertz}, respectively. 
Overall, low-frequency audio has been applied in various contexts despite incomplete evaluation of its privacy protection.

This article systematically addresses the evaluation gap of low-frequency audio and provides further analysis aspects.
The novel contributions are threefold: Our experiments compare two different evaluation scenarios and resampling methods, and analyze the impact of the speaker’s sex and pitch.
Specifically, we evaluate the trade-off between privacy and utility by applying ASR and voice activity detection (VAD) models, to enable comparisons with \cite{liu2024lowfreq, Cohen2019urban}. VAD is an essential pre-processing method for analyzing conversations \cite{parthasarathi2011privacy}, but it may exceed the computational resources available in edge computing scenarios.

The remainder is organized as follows: Section~\ref{sec:method} describes the resampling methods. Our experimental setup is outlined in Section~\ref{sec:setup}. 
Section~\ref{sec:results} presents and discusses the results. Finally, Section~\ref{sec:con} offers conclusions and future research directions.

\section{Resampling audio signals} \label{sec:method}
According to the famous Shannon-Nyquist theorem \cite{shannon1949, nyquist1928}, bandlimited signals with spectral contents up to a maximal frequency $f_{max}$ can be perfectly reconstructed from equally spaced samples with a rate of at least $2f_{max}$, termed the Nyquist rate. 
When a signal is sampled below its Nyquist rate, high-frequency components alias into the low-frequency components.
To prevent these aliasing artifacts, an appropriate low-pass filter is typically applied before sampling.
Utilizing sub-Nyquist sampling for audio coding was addressed in \cite{mishali_sub-nyquist_2011}.

\begin{figure}[t!]
    \centering
    \includegraphics[width=\columnwidth]{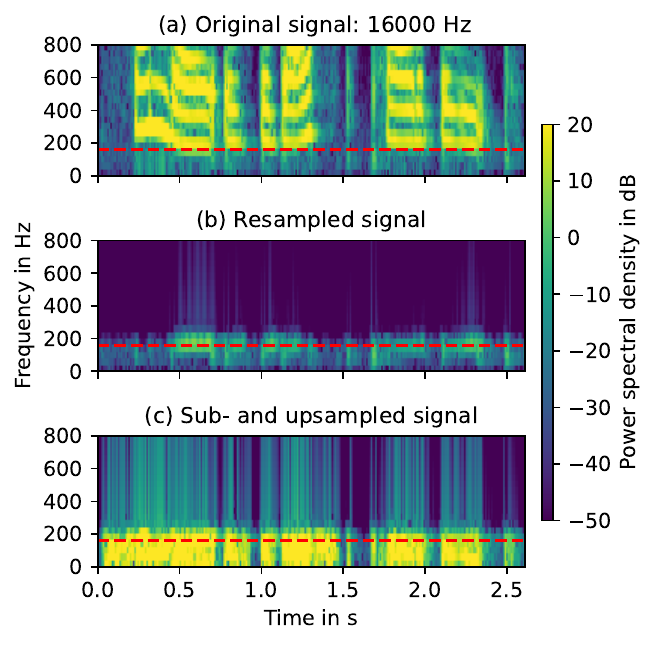}
    \caption{Spectrograms of a female speaker for (a) the original signal at $f_s =$ \SI{16}{\kilo\hertz}, the resampled signal at $f_s =$ \SI{320}{\hertz} with and without anti-aliasing filtering in (b) and (c), respectively. The red line shows the Nyquist frequency of \SI{160}{\hertz}.} 
    \label{fig:resample}
\end{figure}

Recently, 
\cite{hwang_alias-and-separate_2022, lee_improved_2024} utilized aliasing for wideband speech coding at lower bitrates: Speech sampled at \SI{16}{\kilo\hertz} is subsampled by a factor of 2 to result in 
a mixture of the original signal covering 0-\SI{4}{\kilo\hertz} and the spectrally flipped aliasing signal, which corresponds to the original signal covering 8-\SI{4}{\kilo\hertz}.
In general, reducing the sampling rate $f_s$ deteriorates the signal quality.
Therefore, we hypothesize that at low $f_s$, aliased frequencies are useful for reconstructing spoken content, thus compromising speech privacy.
For a female speaker, Figure~\ref{fig:resample} shows the spectrograms of the original signal and the resampled signal at $f_s =$ \SI{320}{\hertz} with and without aliasing. Since the fundamental frequency is in this example above $f_s/2$, the resampled signal contains almost no information, whereas with aliasing, mirrored higher-frequency components are retained.

Conventional sound cards offer a minimum $f_s = \SI{8}{\kilo\hertz}$, corresponding to the standard $f_s$ of most telephony systems. As a result, lower $f_s$ require additional downsampling.
Naive downsampling by subsampling introduces aliasing. To mitigate this, a low-pass filter can be implemented, where its steepness at the cut-off frequency determines the roll-off, i.e., to what extent the filter attenuates frequencies above the cut-off frequency.

In our experiments, the audio signals were resampled using \texttt{torchaudio} (sinc\_interp\_hann) \cite{torchaudio2023}. A preliminary analysis revealed that the resampling method of \texttt{librosa} (soxr\_hq) \cite{mcfee2020librosa} provides better quality, i.e., less roll-off, but results in longer computation times.
We evaluated downsampling rates of 1600, 800, 500, and \SI{320}{\hertz} leading to integer subsampling factors of 10, 20, 32, and 50. We tested both with and without anti-aliasing filtering before downsampling.
Afterwards, all down-sampled signals were upsampled to the original $f_s =$ \SI{16}{\kilo\hertz} to ensure consistency in feature and model parameters.

\section{Experimental setup} \label{sec:setup}
The experimental framework\footnote{Code available at https://github.com/ol-MEGA/ppca.git} \cite{Pohlhausen2025csl} is based on the open-source speech-processing toolkit SpeechBrain \cite{speechbrain}. We conducted the experiments with two Nvidia GeForce RTX 3090 Ti GPU and one Intel Alder Lake CPU. 
All models are based on 80-dimensional log Mel filterbank energies computed on time segments with a window size of \SI{25}{\milli\second} and a hop size of \SI{10}{\milli\second}.
Confidence intervals for the evaluation metrics were calculated with a confidence level $\alpha = 5$~\% using bootstrapping\footnote{https://github.com/luferrer/ConfidenceIntervals}.

\subsection{Evaluation scenarios}\label{sec:attack}
We considered two levels of attackers' knowledge about the resampling method, following the terminology of \cite{Srivastava2020voice_conversion_attacks}: (1) The \textit{ignorant} attacker is not aware that the audio signals were resampled and applies models trained on original data with a sampling rate of \SI{16}{\kilo\hertz} to process low-frequency test data; In contrast, (2) the \textit{informed} attacker knows the resampling method and its exact parameter values. Hence, the informed attacker applies models explicitly trained on low-frequency data.

\subsection{Datasets}\label{sec:Data}
The ASR models were trained on the 360~h train split of LibriSpeech \cite{librispeech} and tested on the test-clean and test-other subsets.
These test signals represent the worst-case scenario for protecting privacy, as background noise or reverberation degrade both attack and utility performance \cite{pohlhausen_2024_iwaenc}.
The VAD models were trained and tested on the train and test split of LibriParty \cite{speechbrain}, respectively.
For both models, audio signals were resampled on-the-fly during inference and after augmentation during training.

\subsection{Automatic speech recognition}\label{sec:ASR}
The state-of-the-art ASR system\footnote{https://huggingface.co/speechbrain/asr-transformer-transformerlm-librispeech} \cite{speechbrain} utilizes a pre-trained transformer-based language model, a pre-trained tokenizer, a transformer acoustic model encoder, and a joint transformer decoder with connectionist temporal classification \cite{CTC}. We employed a batch size of 32 and trained the ASR model for 30 epochs, noting that additional training epochs or larger training datasets may lead to slightly improved performance.
This model has 71.5M parameters and requires approximately \SI{1}{\hour} of training time per epoch, as well as \SI{3}{\hour} for inference on both test sets. 
The transcription performance is assessed with the word error rate (WER), defined as the sum of substitution, insertion, and deletion errors in the ASR output relative to the total number of tokens in the ground-truth transcript. Hence, a high WER is desired to ensure protection of linguistic content.

\subsection{Voice activity detection}\label{sec:VAD}
The VAD model is based on a convolutional recurrent deep neural network (CRDNN) architecture \cite{speechbrain} and performs binary classification using frame-level posterior probabilities. With a total of 109K parameters, the training time per epoch is approximately \SI{10}{\minute}, while inference requires \SI{10}{\second}. The VAD model was trained for 100 epochs.
To evaluate the model's voicing decision, receiver operating characteristic (ROC) curves were calculated, which characterize the trade-off between the true positive rate 
and false positive rate. 
The area under the ROC curve (AUC) was also determined, ranging from 0 to 1.
Moreover, Matthews correlation coefficient (MCC) was calculated, which is derived from the confusion matrix. The MCC ranges from -1 
to 1, 
with 0 indicating random guessing.

\subsection{Analysis of speaker characteristics}\label{sec:pitch}
As shown in Figure~\ref{fig:resample}b, decreasing the Nyquist frequency by downsampling discards higher-frequency harmonics or the fundamental frequency itself. 
Therefore, we hypothesize that the sex of the speakers, or more generally, the pitch of their voices, affects the performance of the ASR and VAD models when processing low-frequency audio. Specifically, we expect that a lower pitch will lead to better transcription results and voicing decisions. 
To investigate this, the YIN\footnote{https://github.com/brentspell/torch-yin} algorithm \cite{yin2002} was applied to estimate the mean pitch for each utterance of the LibriSpeech \cite{librispeech} test-clean set. The pitch range was limited between 80 and \SI{400}{Hz}. 
As the test split of LibriParty \cite{speechbrain} contains noisy speech that can compromise pitch estimation, we divided the \SI{5}{\second}-examples only by sex (F = 37.3~\%, M = 34.2~\%), excluding those with multiple speakers or only noise (28.5~\%).

\section{Results and discussion} \label{sec:results}
This section compares the performance of the ASR models in Section~\ref{sec:asr_results} and of the VAD models in Section~\ref{sec:vad_results}.

\subsection{Automatic speech recognition performance} \label{sec:asr_results}
The following sections report the difference between ignorant and informed ASR models, as well as the impact of sex, pitch, and aliasing on recognition rates.

\subsubsection{Comparing ignorant vs. informed models}
The comparison in Table~\ref{tab:asr_informed_ignorant} highlights that across all $f_s$ and test sets, the total WER of informed models is significantly better than of ignorant models, except for $f_s= \SI{320}{\hertz}$ on the LibriSpeech \cite{librispeech} test-other set, where the difference is negligible, as the WER approaches 100~\%. This advantage of informed models is more pronounced at higher $f_s$, underscoring the importance of retraining or finetuning attacker models for accurate privacy assessments. 
For instance, a sufficient WER~$> 100$~\% of an ignorant model for $f_s= \SI{800}{\hertz}$, which is consistent with \cite{liu2024lowfreq}, strongly overestimates the WER~$= 27.55$~\% of an informed model on clean recordings, thus overestimating the level of privacy protection. 
Moreover, the higher WER~$= 68$~\% \cite{boovaraghavan2024kirigami} on the TIMIT \cite{garofolo1993timit} dataset achieved by a fine-tuned ASR model for $f_s=\SI{1}{\kilo\hertz}$ indicates that complete retraining significantly enhances ASR performance relative to fine-tuning.

\begin{table}[ht]
\centering
\caption{Total WER on the test-clean and test-other sets of LibriSpeech \cite{librispeech}. Ignorant results correspond to an ASR model trained with $f_s= \SI{16}{\kilo\hertz}$, while informed results correspond to ASR models trained with different $f_s$ using resampling. The 95~\% confidence intervals are given in brackets, respectively.}
\begin{tabular}{l@{\hspace{1pt}}c@{\hspace{5.5pt}}c@{\hspace{5.5pt}}c@{\hspace{5.5pt}}c@{}}
\toprule
Sampling & \multicolumn{2}{c}{test-clean}& \multicolumn{2}{c}{test-other} \\ 
rate in Hz & ignorant & informed & ignorant & informed \\ 
\midrule
16000 & - & 2.34 & - & 5.55 \\ 
& & {\scriptsize [1.98, 2.74]} & & {\scriptsize [4.78, 6.50]} \\[1.1ex] 
1600 & 45.99 & 8.90 & 73.10 & 28.14 \\ 
& {\scriptsize [42.83, 49.37]} & {\scriptsize [7.80, 10.19]} & {\scriptsize [70.05, 76.33]} & {\scriptsize [25.52, 31.41]} \\[1.1ex] 
800 & 101.98 & 27.55 & 100.15 & 60.97 \\ 
& {\scriptsize [100.34, 103.86]} & {\scriptsize [24.38, 31.13]} & {\scriptsize [98.56, 101.82]} & {\scriptsize [57.11, 65.15]} \\[1.1ex] 
500 & 95.46 & 52.15 & 95.65 & 85.87 \\ 
& {\scriptsize [94.52, 96.47]} & {\scriptsize [47.55, 56.78]} & {\scriptsize [94.57, 96.72]} & {\scriptsize [82.11, 89.77]} \\[1.1ex] 
320 & 95.97 & 86.91 & 97.15 & 102.37 \\ 
& {\scriptsize [95.20, 96.75]} & {\scriptsize [81.98, 92.65]} & {\scriptsize [96.32, 97.84]} & {\scriptsize [98.96, 105.80]} \\
\bottomrule
\end{tabular}

\label{tab:asr_informed_ignorant}
\end{table}

\subsubsection{Impact of sex and pitch}
\begin{figure}[t]
  \centering
 \begin{subfigure}[b]{\columnwidth}
     \centering
     \caption{ASR results on LibriSpeech test-clean}
     \includegraphics[width=0.99\linewidth]{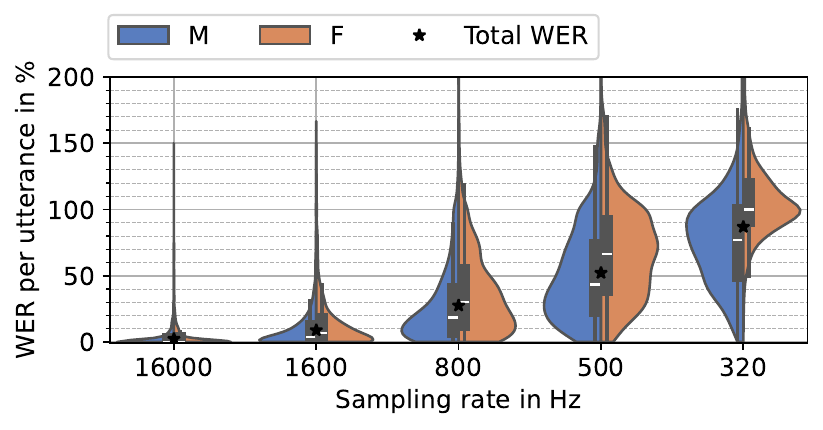}
     \vspace{-0.07cm}
     \label{fig:asr_violin_clean}
 \end{subfigure}
 \begin{subfigure}[b]{\columnwidth}
     \centering
     \caption{ASR results on LibriSpeech test-other}
     \includegraphics[width=0.99\linewidth]{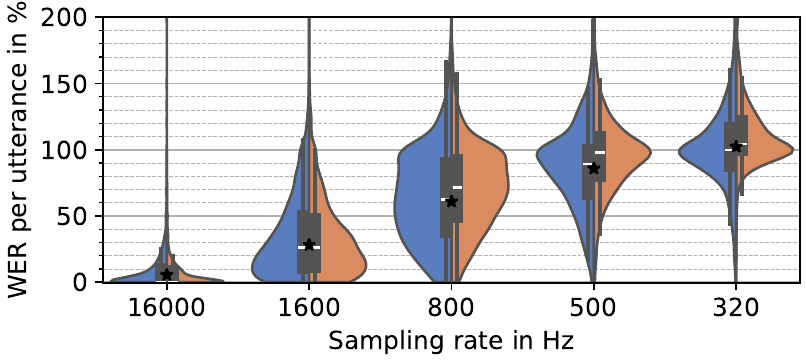}
     \vspace{-0.03cm}
     \label{fig:asr_violin_other}
 \end{subfigure}
\caption{Distribution of the utterance-wise WER on the (a) test-clean and (b) test-other subset of LibriSpeech \cite{librispeech} for ASR models trained with different $f_s$ using resampling. The results are divided by sex, with male speakers on the left of each violin and female speakers on the right. The gray boxes represent the interquartile range. 
The white dash shows the median WER and the star the total WER as in the informed columns of Table~\ref{tab:asr_informed_ignorant}.}
\label{fig:asr_violin}
\end{figure}

For female and male speakers, Figure~\ref{fig:asr_violin} visualizes the distribution of utterance-wise WER with informed ASR models.
As $f_s$ decreases, the impact of sex on WER is more pronounced, with a stronger effect observed on the test-clean compared to the test-other set of LibriSpeech \cite{librispeech}. This latter finding indicates that when ASR performance deteriorates due to challenging recording conditions \cite{librispeech}, the impact of sex is less relevant.
Since the utterance-wise WER is not normally distributed in most cases, we employed Mann-Whitney U tests to evaluate whether the WER for females is stochastically higher than that for males. The tests suggest a significant difference (p-value $<$ 0.01) on both test sets for all $f_s$ except \SI{16}{\kilo\hertz}.

\begin{figure}[b!]
    \centering
    \includegraphics[width=0.94\columnwidth]{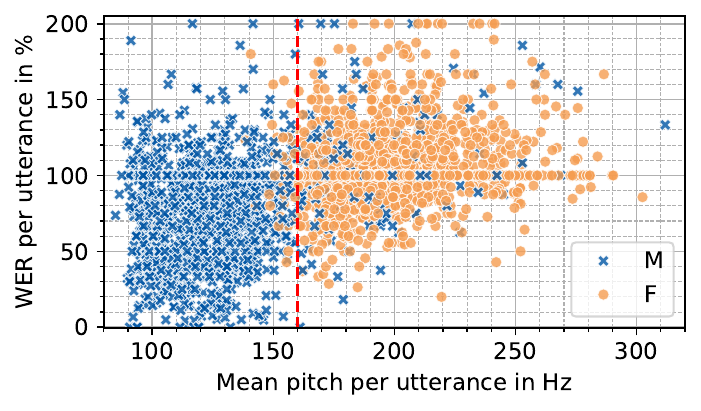}
    \vspace{-0.05cm}
    \caption{Utterance-wise comparison of WER and mean pitch on LibriSpeech \cite{librispeech} test-clean of an ASR model trained with $f_s=$ \SI{320}{\hertz} using resampling. The crosses and dots denote male (M) and female (F) speakers, respectively. The red line shows the Nyquist frequency of \SI{160}{\hertz}.}
    \label{fig:asr_scatter_pitch}
\end{figure}

In addition, Figure~\ref{fig:asr_scatter_pitch} shows for $f_s= \SI{320}{\hertz}$ the pitch dependency of utterance-wise WER on the test-clean set of LibriSpeech \cite{librispeech}. As expected, male speakers exhibit lower pitches and, consequently, lower WER compared to female speakers.
Particularly, 10 male utterances were completely correctly transcribed, but the WER is highly variable for both groups.

\subsubsection{Impact of aliasing}
\begin{figure}[t!]
  \centering
 \includegraphics[width=\linewidth]{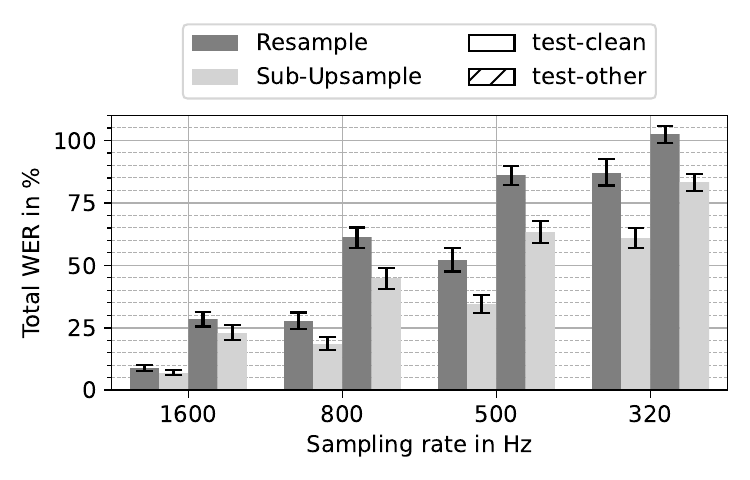}
 \caption{Comparing the total WER on the test-clean and test-other (hatched) set of LibriSpeech \cite{librispeech} for ASR models trained with different $f_s$ using resampled or sub-upsampled signals. Error bars indicate 95~\% confidence intervals.}
\label{fig:asr_resup}
\end{figure}

As expected, the ASR models exploit the additional, aliased information introduced by sub- and upsampling of the audio signals, which significantly improves the total WER for all $f_s$, as shown in Figure~\ref{fig:asr_resup}.
This result urges caution when implementing resampling methods, as aliasing improves the reconstruction of linguistic content.
However, from the perspective of transmitting speech signals, low-frequency audio with aliasing might be of interest because it achieves comparable transcription performance at reduced bandwidth and data storage requirements.

\begin{figure}[b!]
    \centering
    \includegraphics[width=0.9\columnwidth]{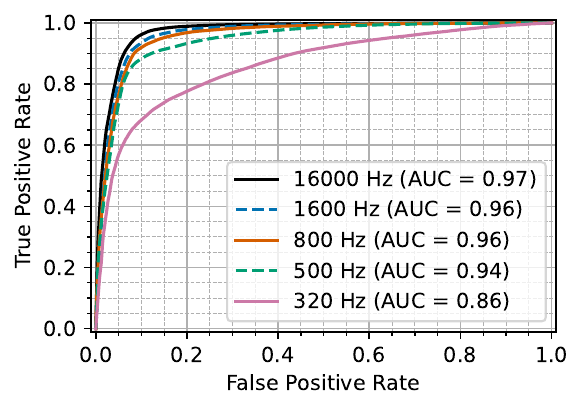}
    \caption{ROC curves on the test split of LibriParty \cite{speechbrain} for VAD models trained with different $f_s$ using resampling.}
    \label{fig:vad_roc}
\end{figure}

\subsection{Voice activity detection performance} \label{sec:vad_results}
This section presents the performance of informed VAD models.
The ROC curves in Figure~\ref{fig:vad_roc} of VAD models trained with resampled signals demonstrate that the AUC remains comparable for $f_s$ of up to \SI{500}{\hertz} to the baseline AUC for $f_s= \SI{16}{\kilo\hertz}$ and deteriorates only for the lowest tested $f_s= \SI{320}{\hertz}$. Consistent with the evaluation of ignorant VAD models \cite{liu2024lowfreq}, the VAD appears robust for low-frequency audio.

\begin{figure}[t!]
    \centering
    \includegraphics[width=\columnwidth]{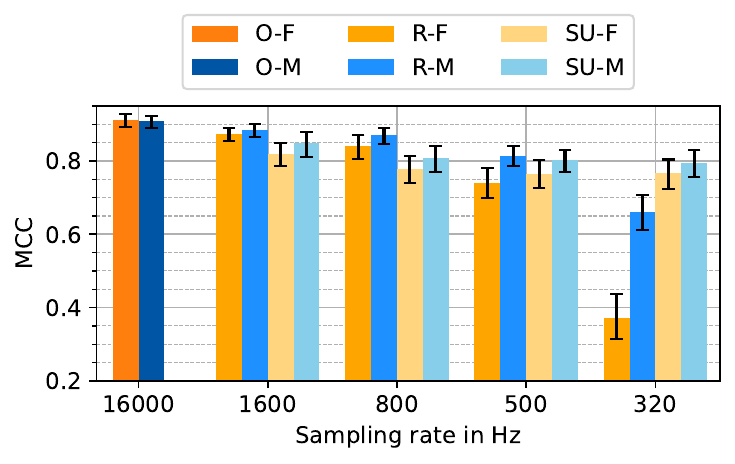}
    \caption{MCC on the test split of LibriParty \cite{speechbrain} for VAD models trained with different sampling rates, comparing the original audio signals (O), resampled (R), and sub-upsampled (SU) signals for female (F) and male (M) speakers. Error bars indicate 95~\% confidence intervals.}
    \label{fig:vad_mcc}
\end{figure}

Figure~\ref{fig:vad_mcc} summarizes the impact of sex and aliasing on the MCC. Similar to the ASR performance shown in Figure~\ref{fig:asr_violin}, differences between female and male speakers are observed, particularly pronounced at lower $f_s$. Specifically for $f_s= \SI{320}{\hertz}$, calculating a sex-independent MCC of 0.51 underestimates the VAD performance for males while overestimating it for females (compare R-F and R-M in Figure~\ref{fig:vad_mcc}).
When aliasing is introduced through sub- and upsampling of the audio signals, the MCC either slightly deteriorates or remains stable for $f_s$ of up to \SI{500}{\hertz}, but strongly improves for $f_s= \SI{320}{\hertz}$. At this lowest $f_s$, aliasing significantly reduces the sex difference in MCC.

\section{Conclusions}\label{sec:con}
This article analyzed the effectiveness of using low-frequency audio to protect spoken content. Simultaneously, the preservation of utility was assessed by applying VAD models. 
Experimental results indicate that informed, retrained ASR models show significantly better transcription performance compared to ignorant, pre-trained ASR models, emphasizing the importance of retraining or finetuning attacker models for accurate privacy assessments. 
For informed models, the total WER exceeds 86~\% only at a sampling rate of \SI{320}{\hertz}, but at the expense of deteriorated VAD performance. This trade-off raises questions about the effectiveness of low-frequency audio as a privacy-protection technique. 
Since low-frequency audio lacks higher-frequency harmonics or the fundamental frequency itself, we observed that the performance of both models was influenced by the speaker's sex and pitch, with both performing better for males. This finding highlights the need for fair and transparent assessments of privacy and utility, as results may be blurred otherwise.
Furthermore, our results demonstrate that both models can exploit aliasing, introduced through sub- and upsampling of the audio signals without anti-aliasing low-pass filtering. These improved performances are relevant for evaluating privacy protection, indicating that the downsampling method should be carefully implemented, not only in audio domains but also in other domains that record data beyond audio. In contrast, resource-limited applications might benefit from aliasing as it reduces bandwidth and data storage requirements.
Future work will explore the impact of using a Mel filterbank with a higher resolution and the combination with different input features.

\section{Acknowledgements}
\ifinterspeechfinal
This work was supported by the Graduation program of Jade University of Applied Sciences (Jade2Pro 2.0).
\fi

\newpage
\bibliographystyle{IEEEtran}
\bibliography{sapstrings, literature}

\end{document}